# Weight-Length Relationships in Gafftopsail Catfish (Bagre marinus) and Hardhead Catfish (Ariopsis felis) in Louisiana Waters

Joshua Courtney,[1] Taylor Klinkmann,[2] Joseph Toraño,[2] and Michael Courtney[2]

[1]BTG Research, P.O. Box 62541, Colorado Springs, CO, 80962

[2]United States Air Force Academy, 2354 Fairchild Drive, USAF Academy, CO, 80840

**Abstract**

In spite of the abundance and commercial importance of these two species, there is little published weight-length data for the gafftopsail catfish (*Bagre marinus)* and hardhead catfish *(Ariopsis felis).* For this study 84 catfish were caught (hook and line) from the Calcasieu Estuary in Southwest Louisiana near the Gulf of Mexico and estuaries and near shore waters close to bayou Lafourche. Using least squares regression, best fit curves were determined for weight (W) vs. total length (L) relationships in gafftopsail catfish, $W(L) = 1000(L/484.73)^{3.2660}$, while the best-fit equation for the hardhead was $W(L) = 1000(L/469.53)^{3.0188}$, where W is the weight in grams and L is the total length in millimeters. Results showed that, when compared to gafftopsail and hardhead catfish caught in Florida, Louisiana gafftopsail catfish tend to weigh less at similar lengths; whereas, the Louisiana hardhead catfish tend to weigh about the same. Results also show that the total length (TL) can be related to the fork length (FL) as TL = 1.184 FL in gafftopsail catfish with $R^2 = 0.9922$ and TL = 1.140 FL with $R^2 = 0.9667$ in hardhead catfish.

## Introduction

Published weight-length relationships for the gafftopsail catfish (*Bagre marinus*) and hardhead catfish (*Ariopsis felis*) are scarce.[1] This is unfortunate, as both of these fish have significant commercial value. From 1987 to 2001, 1.04 million pounds of these catfish were harvested for commercial purposes in the 5-county area of Volusia, Brevard, Indian River, St. Lucie, and Martin in Florida.[2] Overall, this harvest was valued at $777,497. From 1959 to 1967, an estimated total of 3.67 million pounds were harvested commercially in the waters of Texas, Louisiana, Mississippi, and Florida.[3]

The gafftopsail catfish is most commonly found in the western central Atlantic Ocean, the Gulf of Mexico, and the Caribbean Sea.[4] Its spines are venomous and cause painful wounds, which often discourages fishermen from catching them. Its diet consists chiefly of crustaceans: crabs, shrimp, and prawns. To establish a weight-length curve for the gafftopsail, forty-one fish were caught and measured from inshore waters in the bayou Lafourche area including near shore waters of the Gulf of Mexico.

The hardhead catfish is most commonly found near the shores of the Western Atlantic Ocean, particularly in the Gulf of Mexico, the Florida Keys, and the southeast coast of the US.[1] It gets its name from the signature bony plate extending from its eyes to the dorsal fin. Its diet consists of crustaceans, algae, sea grasses, and a variety of small fish. Both species are particularly abundant in the two study areas, inshore waters of the Calcasieu estuary and adjacent near shore waters and inshore waters near bayou Lafourche and adjacent near shore waters.[3]

## Method

The catfish used in this study were caught via hook and line by sport anglers. Creel surveys were performed over a three week period from late May to mid-June, 2011. In each of two study areas, anglers were asked for permission to weigh and measure fish from their ice chests. One study area was the Calcasieu Estuary in

# Weight-Length Relationships in Gafftopsail Catfish (Bagre marinus) and Hardhead Catfish (Ariopsis felis) in Louisiana Waters

southwestern Louisiana centered near +29° 56' 45.64 N", -93° 18' 12.84" W and extending approximately 20 km north and south and 10 km east and west from this point, including 2 km into the Gulf of Mexico near Calcasieu pass. Specifically, creel surveys were performed at boat launches at Calcasieu Point, Hebert's Marina, and the Cameron Jetties (referred to henceforth as Calcasieu). The other study area was in Lafourche Parish, Louisiana, centered near +29° 15' 10.28"N, -90° 12' 44.24" W and extending approximately 25 km north and south and 15 km east and west and included Bayou Lafourche from Golden Meadow, south to the estuaries including Barataria Bay and Timbalier Bay and up to 2 km south of the Gulf Coast (referred to henceforth as th Lafourche area). Specifically, creel surveys were performed at Bobby Lynn's Marina and the Port Fourchon public boat launch.

Each fish was weighed to the nearest .01 kg. The fork length (length from the front of the mouth to the vertex of the fork in the tail) and total length (length from the front of the mouth to the longest point of the tail) of each fish were measured to the nearest 3.2 mm (1/8 inch). The traditional weight length relationship in fish has the form $W = aL^b$, where $W$ is the weight (grams), $L$ is the total length (millimeters), and $a$ and $b$ are values found by fitting an equation of this form to measurement data for a given species of fish. The power $b$ is close to 3 for most species of fish. To establish a weight-length curve for each species, linear least-squares (LLS) regression (determined by $\log(W)=\log(a) + b \log(L)$) was used to determine the parameters of the power law best-fit equation, $W(L) = aL^b$. Non-linear least-squares (NLLS) regression was used to determine a customized best-fit equation, also $W(L) = aL^b$, and an improved model, $W(L)= 1000(L/L_1)^b$. This model has been shown to yield smaller uncertainties in the parameters $L_1$, which is the typical length of the fish weighing 1 kg, and the exponent $b$.[5,6,7] If desired, the equivalent parameter $a$ from the traditional model can be computed as $a = 1000L_1^{-b}$.

## Results

Figure 1 shows the weight-length relationship of gafftopsail catfish. The data includes 41 fish. The largest sample is 2.96 kg, 679 mm, and the smallest sample is 0.1 kg, 235 mm. The Florida weight vs. length curves lie close to each other and slightly above the best fit curves for the Louisiana data in the present study. The Louisiana best fit curves are close to each other for the LLS and NLLS fits, but the parameters are not exactly equivalent as shown in Table 1, and the curves do not exactly overlap. The parameters obtained by LLS and NLLS methods differ significantly from each other, but the parameters obtained by the traditional and improved models are nearly equivalent when obtained by NLLS.

Figure 2 shows the weight-length relationship for hardhead catfish along with best fit curves and weight-length relationships plotted for male and female hardhead catfish in a Florida study. The data included 43 fish from the Calcasieu estuary and estuaries near Bayou Lafourche, Louisiana, the largest sample being 866 g, 433 mm, and the smallest sample being 101 g, 241 mm. The curves are all close to each other and demonstrate substantial overlap.

# Weight-Length Relationships in Gafftopsail Catfish (Bagre marinus) and Hardhead Catfish (Ariopsis felis) in Louisiana Waters

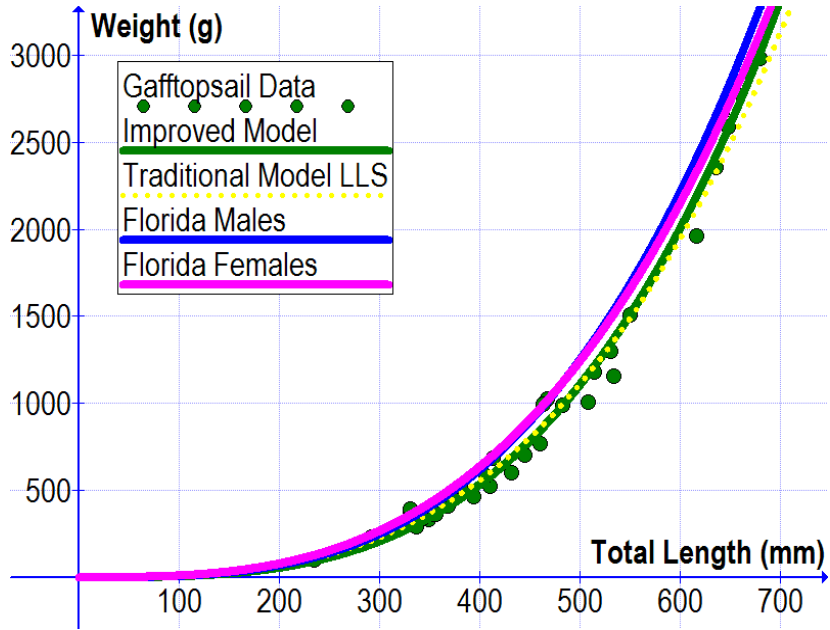

*Figure 1: Weight-length data for gafftopsail catfish with best fit curves obtained using non-linear least squares for the improved model $W(L) = 1000(L/L_1)^b$ and linear least squares for the traditional model. Curves for male and female weight-length relationships from Florida are shown for comparison.[8] The relationship between fork length (FL) and total length (TL), TL = 1.184 FL, ($R^2$ = 0.9922) obtained from the present study is used to convert the Florida equations from FL to TL.*

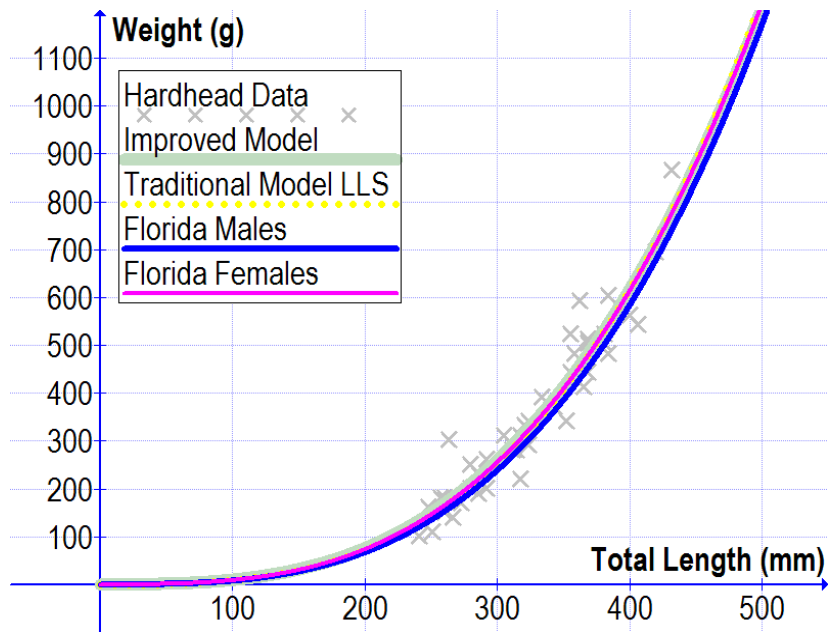

*Figure 2: Weight-length data for hardhead catfish with best fit curves obtained using non-linear least squares for the improved model $W(L) = 1000(L/L_1)^b$ and linear least squares for the traditional model. Curves for male and female weight-length relationships from Florida are shown for comparison.[8] The relationship between fork length (FL) and total length (TL), TL = 1.140 FL, ($R^2$ = 0.9667) obtained from the present study is used to convert the Florida equations from FL to TL.*

# Weight-Length Relationships in Gafftopsail Catfish (Bagre marinus) and Hardhead Catfish (Ariopsis felis) in Louisiana Waters

| | Parameter | Value | Uncertainty |
|---|---|---|---|
| Gafftopsail Catfish | | | |
| $W = aL^b$ (LLS) | a | 5.2665E-06 | 34.5% |
| $R^2 = 0.9865$ | b | 3.0843 | 1.8% |
| $W = aL^b$ (NLLS) | a | 1.695E-06 | 42.1% |
| $R^2 = 0.9902$ | b | 3.2660 | 2.0% |
| $W = (L/L_1)^b$ | $L_1$ | 484.73 mm | 0.5% |
| $R^2 = 0.9902$ | b | 3.2660 | 2.0% |
| | $a=1000L_1^{-b}$ | 1.695E-06 | 1.8% |
| Hardhead Catfish | | | |
| $W = aL^b$ (LLS) | a | 5.719E-06 | 90.6% |
| $R^2 = 0.9020$ | b | 3.087 | 5.1% |
| $W = aL^b$ (NLLS) | a | 8.604E-06 | 94.6% |
| $R^2 = 0.9133$ | b | 3.0188 | 5.2% |
| $W = (L/L_1)^b$ (NLLS) | $L_1$ | 469.53 mm | 1.5% |
| $R^2 = 0.9133$ | b | 3.0188 | 5.2% |
| | $a=1000L_1^{-b}$ | 8.606E-6 | 4.5% |

*Table 1: Parameters for best-fit equations for weight-length relationships in gafftopsail and hardhead catfish.*

## Discussion

Both the improved equation and traditional power law equation using non-linear least-squares provided better best-fit curves than the power law equation solved with linear least-squares using the log-log technique. For the gafftopsail, the $R^2$ of the improved and traditional equations was 0.9902; whereas, the $R^2$ of the power law equation using LLS was 0.9865. For the hardhead, the $R^2$ of the improved and traditional equation when solved via NLLS was 0.9133; whereas, the $R^2$ of the power law equation was 0.9120. The uncertainty in the parameter *a* is much smaller (by a factor of 15-20) using the improved model, fitting for $L_1$, and then solving for *a*. As a result, the data shows the improved model to be the best in determining the weight-length relationships in these two catfish. However, in spite of the advantage of reduced uncertainty in the parameters, the curves generated by the best fit models do not differ greatly.

One of the few other studies done on the weight-length relationship of gafftopsail and hardhead catfish was published by the Florida Fish and Wildlife Conservation Committee in 2010.[8] Using the best-fit equation $W(L) = aL^b$, the a and b parameters of the gafftopsail catfish were reported to be 0.000401 and 3.143 respectively for males and 0.000585 and 3.007 respectively for females. For the hardhead catfish, the a and b parameters are 0.000361 and 3.116 respectively for males and 0.000448 and 3.053 respectively for females.[8] As Figures 1 and 2 show, both male and female gafftopsail catfish in Florida tended to be heavier at a given length than their unsexed counterparts in Louisiana. In contrast, the hardhead catfish tended to be closer in weight between the two states at a given length.

# Weight-Length Relationships in Gafftopsail Catfish (Bagre marinus) and Hardhead Catfish (Ariopsis felis) in Louisiana Waters

## Acknowledgements

This work was funded by BTG Research (www.btgresearch.org). The authors also appreciate the advice, hospitality, expertise, and facilities shared by Matthew Courtney.